\documentclass[twocolumn,twocolappendix]{aastex701}
\usepackage{amsmath}
\usepackage{array}

\begin{document}

\title{
Cosmic-Ray Spectra and Metal Budget Regulated by the Galactic Wind}

\author[sname='Fukumoto']{Yusaku Fukumoto}
\affiliation{Institute for Cosmic Ray Research, The University of Tokyo, 
5-1-5 Kashiwanoha, Kashiwa, Chiba 277-8582, Japan}
\email{fukumoto@icrr.u-tokyo.ac.jp}

\author[0000-0001-9064-160X,sname='Asano']{Katsuaki Asano}
\affiliation{Institute for Cosmic Ray Research, The University of Tokyo, 
5-1-5 Kashiwanoha, Kashiwa, Chiba 277-8582, Japan}
\email[show]{asanok@icrr.u-tokyo.ac.jp}

\author[0000-0003-3383-2279,sname='Shimoda']{Jiro Shimoda}
\affiliation{Institute for Cosmic Ray Research, The University of Tokyo, 
5-1-5 Kashiwanoha, Kashiwa, Chiba 277-8582, Japan}
\email{jshimoda@icrr.u-tokyo.ac.jp}

\begin{abstract}
We study the advection effect of the galactic wind on the local cosmic-ray (CR) spectra. The spectral hardening from a few hundred GV and softening from a few TV are reproduced by a velocity profile with a maximum velocity of $\sim 700~\mbox{km}~ \mbox{s}^{-1}$
without introducing a break in the power-law dependence of the diffusion coefficient.
Additionally, we find that a hard CR spectrum below $\sim$ TV with an index of $\sim 2$ at an altitude of $\sim 3$-5 kpc from the Galactic disk.
This hard spectrum is favorable for the gamma-ray spectrum of the Fermi bubbles.
With the obtained CR fluxes, we discuss the matter circulation in our Galaxy with the wind. While the wind has an essential role in maintaining the metal abundance in the disk, the production rate of beryllium, which originates from CR spallation, is so low that the ratio Be/O in the halo should be larger than that in the disk gas.
\end{abstract}



\section{Introduction}
\label{sec:intro}

The observed cosmic-ray (CR) spectra show a hardening above a few hundred GV and a softening around a multi-TV region \citep[e.g.][]{2011Sci...332...69A,2017PhRvL.119y1101A,2019SciA....5.3793A,adriani22,2023PhRvL.130q1002A}.
Several models have been proposed for this feature, e.g., local source contribution 
\citep{2012ApJ...752...68V,2012MNRAS.421.1209T,2017PhRvD..96b3006L,2021ApJ...917...61K,2025arXiv251207239S},
modification of the momentum dependence of the diffusion coefficient \citep{2012PhRvL.109f1101B,2013JCAP...07..001A,2018PhRvL.121b1102E}, reacceleration \citep{2014A&A...567A..33T}, spatial inhomogeneity of the diffusion coefficient \citep{2012ApJ...752L..13T,2018PhRvD..97f3008G},
and so on.

Another attractive mechanism for the spectral feature is the advection effect of the galactic wind starting from a high altitude, proposed by \citet{2017PhRvD..95b3001T}.
This opens a new possibility linking CR physics and Galaxy evolution.
The galactic wind is an indispensable component to describe the evolution of star formation and metallicity in our Galaxy \citep{tumlinson17,andrews17,shimoda22a,shimoda24,griffith25,johnson25}.
The star formation rate in our Galaxy has been stable for $\sim 10$ Gyr \citep{haywood16,gallart19,fantin19,mor19},
which suggests that the physical properties of the Galactic disk remain almost unchanged.
The gas amount in the disk should be maintained by the balance between the gas consumption due to star formation, wind mass loss, and accretion.
As the wind is driven by star formation, which is triggered by the gas accretion, the gas budget in the disk may be self-regulated.
In addition, the removal of metal by the wind is essential for the present metallicity in the disk.
The metal-polluted gases extended at a scale of $\sim100$~kpc in other galaxies \citep{tumlinson17} also indicate the role of the wind in the metal abundance.

The Fermi bubble \citep{su10,ackermann14} can be a signature of the interaction between Galactic CRs and wind. \citet{shimoda_asano24} show that $\pi^0$-decay gamma rays are emitted from the high-density region formed by the fallback winds.
However, the obtained gamma-ray spectrum is softer than the observed one.
To reproduce the observed spectrum, a more sophisticated treatment may be needed for CR propagation.

While the modification of the diffusion coefficient is an attractive model for the CR spectral feature in the context of plasma physics \citep[e.g., due to CR-induced waves, see][]{2012PhRvL.109f1101B}, the macroscopic phenomenon of the galactic wind should also affect the CR propagation. It is worth estimating the effect of the Galactic wind on the CR spectra, which can make the CR measurement a tool for exploring Galaxy evolution.

In this paper, following the model of \citet{2017PhRvD..95b3001T}, in which the measured CR spectra are not fitted, we demonstrate that the advection effect by the Galactic disk can reproduce the measured spectral hardening and softening.
As the spectra measured with different instruments differ slightly from each other, we primarily focus on the CALET data in this paper, as similarly studied in \citet{2022ApJ...926....5A}.
Then, based on the obtained results, we discuss the circulation of gas and metal in our Galaxy.
This paper is organized as follows. Our simplified one-dimensional model is explained in section \ref{sec:model}.
The results for the CR spectra are shown in section \ref{sec:spectra}. Section \ref{sec:gas-circ} is devoted to discussing the gas and metal budgets in the disk with the galactic wind. Our conclusions are summarized in section \ref{sec:summary}

\section{Model Setup}
\label{sec:model}

The distribution of CRs escaping from a uniform disk can be approximated as a cylindrical one near the disk. At a height significantly higher than the disk size, CRs spherically propagate. 
As a simple model, we consider a one-dimensional diverging tube perpendicular to the Galactic disk. The cross section of the tube is almost constant for a lower altitude $z$, but expands like a sphere at a significantly high altitude as $\propto z^2$.
In our model, CRs propagate along the tube, neglecting the horizontal diffusion. Assuming an isotropic momentum distribution, the evolution of the CR spectral momentum density $N_i(z,p,t)$ for a nuclide $i$ of mass $m_i$ and charge $Z_i$ is written as 
\begin{eqnarray}
&&\frac{\partial N_i}{\partial t}
= 
 \nabla \cdot \left[ D(p)\, \frac{\partial}{\partial z}  N_i\right] - \nabla \cdot \left[V(z)\, N_i\right] \nonumber \\
&& 
+ \frac{\partial}{\partial p}\left[ \frac{p}{3}\left(\nabla \cdot  V(z) \right) N_i\right] 
+ \frac{\partial}{\partial p}\left[ \dot{p}_i( z,p)\, N_i\right] \nonumber \\
&&+ Q_i(z,p)
- \frac{N_i}{\tau_{i,{\mathrm{col}}}(z,p)}
- \frac{N_i}{\tau_{i,{\mathrm{dec}}}(p)},
\label{eq:propagation}
\end{eqnarray}
where the right-hand terms express the vertical diffusion, advection by the wind, adiabatic work, momentum loss, injection, spallation, and decay, respectively.
Here, the expansion of the tube is taken into account in the divergence operator as
\begin{eqnarray}
\nabla \cdot \equiv \frac{1}{{\cal A}(z)} \frac{\partial}{\partial z}{\cal A}(z),
\end{eqnarray}
where
\begin{eqnarray}
{\cal A}(z) = z^2 + z_{\rm tube}^2.
\end{eqnarray}
In this paper, the parameter $z_{\rm tube}$ is chosen as 10 kpc, far above which CRs spherically propagate.

\begin{figure}
\includegraphics[width=\linewidth]{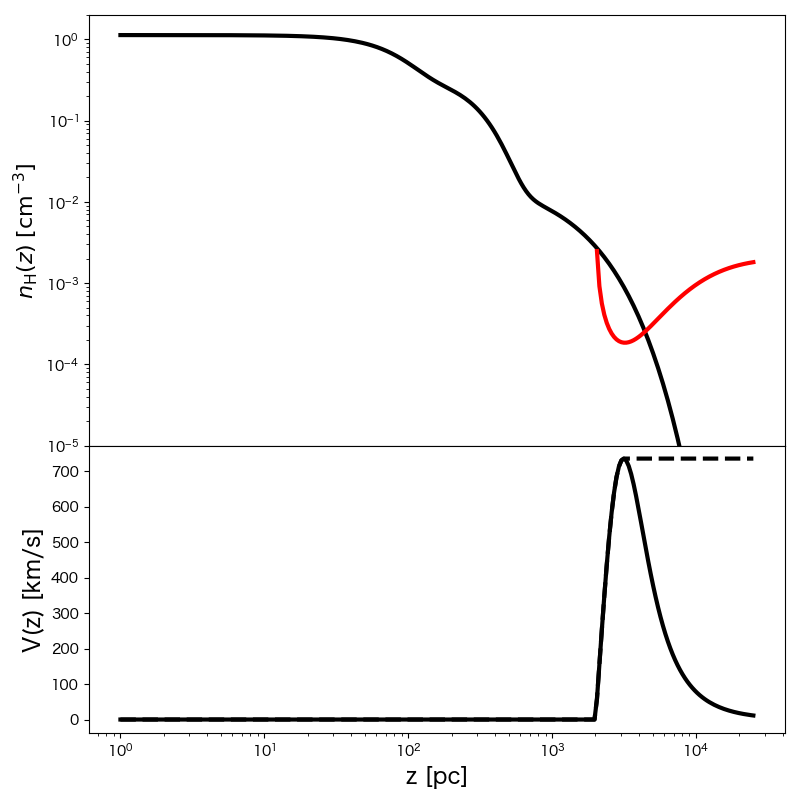}
\caption{The total hydrogen density at $R=8.5$ kpc (upper) and wind velocity profile (lower) in our model.
The red line in the upper panel is the wind density profile with the mass loss rate of $7.26\times 10^{-3}M_\odot \mbox{kpc}^{-2}\mbox{yr}^{-1}$ ($5.13 M_\odot \mbox{yr}^{-1}$) implied from the quasi-steady gas circulation.
The dashed line in the lower panel is the test case with the constant velocity.}
\label{fig:profile}
\end{figure}

To estimate the momentum loss and spallation due to the interaction with the background gas, we model the altitude distribution of hydrogen gases above the Sun, which is at the radius $R=8.5\,\mathrm{kpc}$ from the Galactic center.
At $z=0$, the molecular hydrogen density is $n_1=0.385 ~\mbox{cm}^{-3}$ \citep{1988ApJ...324..248B} and the hydrogen atom density is $n_2=0.34 ~\mbox{cm}^{-3}$ \citep{1976ApJ...208..346G}.
Following \citet{1998ApJ...509..212S}, the molecular and atomic densities are expressed as
\begin{eqnarray}
n_{\mathrm{H}_2}(z) &= n_1
\exp\left[-\log(2)\left(\frac{z}{70\,\mathrm{pc}}\right)^2\right], \\
n_{\rm HI}(z) &= n_2
\exp\left[-\log(2)\left(\frac{z}{250\,\mathrm{pc}}\right)^2\right],
\end{eqnarray}
respectively.
The ionized hydrogen density is given by
\begin{eqnarray}
n_{\mathrm{HII}}(z) &=& n_3
\exp\left[-\frac{z}{1\,\mathrm{kpc}}
-\left(\frac{R}{20\,\mathrm{kpc}}\right)^2\right] \nonumber \\
&&\quad + n_4
\exp\left[-\frac{z}{150\,\mathrm{pc}}
-\left(\frac{R}{2\,\mathrm{kpc}}-2\right)^2\right],
\end{eqnarray}
with $n_3=0.025 ~\mbox{cm}^{-3}$ and $n_4=0.2 ~\mbox{cm}^{-3}$ \citep{1998ApJ...509..212S},
which is consistent with the recent model of \citet{2017ApJ...835...29Y}. The total hydrogen density is
\begin{eqnarray}
n_{\mathrm{H}}(z) = 2n_{\mathrm{H}_2}(z)
+ n_{\mathrm{HI}}(z) + n_{\mathrm{HII}}(z),
\end{eqnarray}
which is plotted in the upper panel of Figure \ref{fig:profile}.

As primary CRs, we consider nuclides H, C, N, and O in our calculation.
Those CR nuclei are injected from supernovae, whose distribution may be proportional to the molecular density.
With the $n_{\mathrm{H}_2}$ distribution, we model the primary CR injection rate as a broken power-law of momentum:
\begin{eqnarray}
Q_{i}(z,p) &=& Q_0 \frac{n_{\mathrm{H}_2}(z)}{n_1} \nonumber \\
&&\times
\begin{cases}
\left(\frac{{\cal R}}{1\,\mathrm{GV}}\right)^{s_1} & ({\cal R}<{\cal R}_{\rm br})\\
\left(\frac{{\cal R}_{\rm br}}{1\,\mathrm{GV}}\right)^{s_1} \left(\frac{{\cal R}}{{\cal R}_{\rm br}}\right)^{s_2} & ({\cal R}>{\cal R}_{\rm br}),
\end{cases}
\end{eqnarray}
where
\begin{equation}
{\cal R} = \frac{p c}{Z_i e},
\end{equation}
is the rigidity of the CR nuclide $i$.

Collisions of C, N, and O CRs with the background hydrogen produce secondary CRs of $i=^6$Li,
$^7$Li, $^7$Be, $^9$Be, $^{10}$Be, $^{10}$B, and $^{11}$B. For those secondary CRs, assuming the conservation of energy per nuclei, the injection rate is calculated with
\begin{eqnarray}
Q_{i}(z,p) = \sum_j \sigma_{ji} (p') v_j(p') N_j(z,p') n_\mathrm{H}(z),
\label{eq:sec}
\end{eqnarray}
where $p'=(m_j/m_i)p$, and $j=$C, N, and O.
Given the energy $E=\sqrt{p^2c^2 + m_j^2c^4}$, the velocity of the primary CR is written as $v_j=pc^2/E$.
The cross sections $\sigma_{ji}$ are obtained from those in the CR propagation code DRAGON2 \citep{2018JCAP...07..006E}.
The collision timescale for spallation is written as
\begin{eqnarray}
\tau_{i,{\mathrm{col}}}(z,p)=\frac{1}{\sum_j \sigma_{ij} (p) v_i(p) n_\mathrm{H}(z)}.
\end{eqnarray}

While an atomic $^{7}$Be decays via electron capture, a CR $^{7}$Be can be treated as a stable particle.
In the secondary CR nuclides we calculate, only $^{10}$Be nuclides are unstable (decay into $^{10}$B); the decay time is
\begin{eqnarray}
\tau_{{\rm Be},{\mathrm{dec}}}(p)=0.961 \frac{E}{10 m_{\rm p}c^2}~\mbox{Myr}.
\end{eqnarray}

For low-energy CRs, the energy loss via interaction with the background gases is efficient.
We use the formulae in \citet{1998ApJ...509..212S} for the energy loss. The energy loss rate due to the Coulomb scattering with ionized gases \citep{1994A&A...286..983M} is
\begin{eqnarray}
\left( \frac{dE_i}{dt} \right)_{\rm sc}
=
4\pi r_{\rm e}^2 m_{\rm e} c^3 Z_i^2 n_{\mathrm{HII}} \ln\Lambda
\frac{\beta^2}{x_m^3 + \beta^3},
\end{eqnarray}
with
\begin{equation}
x_m
=
\left[
\frac{3\sqrt{\pi}}{4}
\right]^{1/3}
\left(
\frac{2kT_{\rm e}}{m_{\rm e} c^2}
\right)^{1/2},
\end{equation}
and the Coulomb logarithm,
\begin{equation}
\ln\Lambda
=
\frac{1}{2}
\ln\!\left(
\frac{m_{\rm e}^2 c^4}{\pi r_{\rm e} \hbar^2 c^2 n_{\mathrm{HII}}}
\frac{m_i \gamma^2 \beta^4}{m_i + 2\gamma m_{\rm e}}
\right),
\end{equation}
where $\beta\equiv v/c$, $\gamma=1/\sqrt{1-\beta^2}$, and $r_{\rm e}\equiv e^2/(m_{\rm e} c^2)$ is the classical electron radius. In our calculation, we adopt $10^4$K for the electron temperature $T_{\rm e}$.
The energy loss rate due to the ionization of neutral gases
\citep{1994A&A...286..983M} is
\begin{eqnarray}
\left( \frac{dE_i}{dt} \right)_{\rm ion}=
2\pi r_{\rm e}^2 m_{\rm e} c^3 Z_i^2
\frac{n_{\rm HI}}{\beta}
\sum_{s={\rm H,He}}
f_s B_s,
\end{eqnarray}
for $\beta>1.4e^2/(\hbar c)$,
where
\begin{eqnarray}
B_s
=
\ln\!\left(
\frac{2 m_{\rm e} c^2 \beta^2 \gamma^2 Q_{\max}}{I_s^2}
\right)
- 2\beta^2,
\end{eqnarray}
with
\begin{eqnarray}
Q_{\rm max}
=
\frac{2 m_{\rm e} c^2 \beta^2 \gamma^2}
{1 + (2\gamma m_{\rm e}/m_i)}.
\end{eqnarray}
The number fraction $f_s$ of nuclide $s$ is assumed to be the solar abundance (see Table \ref{table:fi}).
For the effective ionization potential $I_s$, we use $I_{\rm H}=19\,\mathrm{eV}$, and $I_{\rm He}=44\,\mathrm{eV}$.
For equation (\ref{eq:propagation}), the momentum loss rate is written as
\begin{eqnarray}
\dot{p}_i
&=&
\frac{E}
{pc^2} \left( \left( \frac{dE_i}{dt} \right)_{\rm sc}+ \left( \frac{dE_i}{dt} \right)_{\rm ion} \right) \\
&=&
\frac{1}
{v} \left( \left( \frac{dE_i}{dt} \right)_{\rm sc}+ \left( \frac{dE_i}{dt} \right)_{\rm ion} \right).
\label{eq:pdot}
\end{eqnarray}

\begin{table}
	\centering
	\caption{Nuclear composition ratio in the Galactic disk \citep{2009M&PSA..72.5154L}}
	\begin{tabular}{cc}
	\hline
    \hline
	$i$ &$f_i$ \\
	\hline
	H & 1.0  \\
	He & $9.7\times 10^{-2}$\\
        $^6$Li & $1.62\times 10^{-10}$\\
        $^7$Li & $1.98\times 10^{-9}$\\
        $^9$Be & $2.36\times 10^{-11}$\\
        $^{10}$B & $1.43\times 10^{-10}$\\
        $^{11}$B & $5.83\times 10^{-10}$\\
        C & $2.78\times 10^{-4}$\\
        N & $8.22\times 10^{-5}$\\
        O & $6.08\times 10^{-4}$\\
        Fe & $3.27\times 10^{-5}$\\
	\hline
    \label{table:fi}
	\end{tabular}
\end{table}

\section{Cosmic-Ray Spectra}
\label{sec:spectra}

We numerically solve spatially one-dimensional equation (\ref{eq:propagation})
to obtain the temporal evolution of the momentum distribution $N_i(z,p,t)$, which becomes steady state after $\sim1$ Gyr with temporarily constant injection rates and $z_{\rm tube}=10$ kpc.
To resolve the wind profile, the spatial grid is set as $z_0=0$ pc, $z_1=30$ pc, and $z_k=(2.5 k^2+27.5 k+5)$ pc, where the outermost grid is $k=95$, corresponding to $\simeq 25$ kpc.
For the momentum space, we logarithmically divide it into 300 grids.

\begin{table*}
	\centering
	\caption{Injection parameters and resultant surface density of the energy injection}
	\begin{tabular}{cccccc}
	\hline
    \hline
	& $Q_0$ & ${\cal R}_{\rm br}$ & $s_1$ & $s_2$ & $dE/dtdS$ \\
    	& $[\mbox{cm}^{-3}  \mbox{yr}^{-1} (\mbox{GeV}/c)^{-1}]$& $[\mbox{GV}]$ &  & & $[\mbox{erg}~ \mbox{kpc}^{-2} \mbox{yr}^{-1}]$ \\
	\hline
	H & $1.57\times10^{-16}$ & 1.5 & $-1.3$ & $-2.48$ & $3.10\times10^{45}$ \\
	C & $9.96\times10^{-20}$ & 1.5 & $-1.0$ & $-2.37$ & $1.25\times10^{44}$\\
        N & $1.66\times10^{-20}$ & 2.5 & $-1.0$ & $-2.50$ & $3.96\times10^{43}$ \\
        O & $7.80\times10^{-20}$ & 1.5 & $-1.0$ & $-2.37$ & $1.77\times10^{44}$\\
	\hline
    \label{tab:inj}
	\end{tabular}
\end{table*}

First, we determine the parameters for the diffusion coefficient $D(p)$, wind velocity profile $V(z)$, and CR injection (see Table \ref{tab:inj}), by reproducing the CR spectra for C, N, O, and B at $z=0$.
We especially focus on the CALET data for those spectra.
We use a simple power-law for the momentum dependence of the diffusion coefficient as
\begin{eqnarray}
D(p) = D_0 \left(\frac{{\cal R}}{1\,\mathrm{GV}}\right)^\delta.
\end{eqnarray}
\citet{2017PhRvD..95b3001T} demonstrate that a spectral bump can be produced by a wind starting from a high altitude, rapidly accelerating and turning into deceleration.
Similarly to this model, using the hyperbolic tangent function, we assume
\begin{eqnarray}
V(z) =
\begin{cases}
0 & (0<z<z_{\rm w})\\
V_0  \frac{\tanh{\left(\frac{z - z_{\rm w}}{d}\right)}}{\tanh(1)}\frac{2}{1+\left(\frac{z-z_{\rm w}}{d}\right)^\alpha} & (z>z_{\rm w}),
\end{cases}
\label{eq:wind}
\end{eqnarray}
where $V_0$ roughly corresponds to the maximum velocity of the wind. From $z=z_{\rm w}$, the wind rapidly accelerates as far as $z\sim z_{\rm w}+d$, and turns into deceleration, which is adjusted by the index $\alpha$.
As the actual velocity field is complex and variable, this velocity profile should be regarded as the effective profile obtained from the time average of the velocity field.

Note that our purpose is to find approximate wind conditions to reproduce the spectral bumps around ${\cal R}\sim 1$ TV, rather than the best-fit parameters.
The parameters we propose in this paper are $D_0=1.22\times 10^{28} \mbox{cm}^2 \mbox{s}^{-1}$, $\delta=0.40$, $V_0 = 700~\mbox{km}~ \mbox{s}^{-1}$, $z_{\rm w}=2$ kpc, $d=1.5$ kpc, and $\alpha=1.86$. The wind profile is shown in the lower panel of Figure \ref{fig:profile}.
The maximum velocity is $756 \mbox{km}~\mbox{s}^{-1}$.
The injection parameters for primary CRs are summarized in Table \ref{tab:inj}.
Given the calculated $N_i$, the CR flux is calculated as
\begin{eqnarray}
\frac{d {\cal N}_i}{dt dS d\Omega d(E_{\rm kin}/A)} &=& \frac{v_i}{4\pi} \frac{E}{p c^2} A N_i(z,p) \\
&=& \frac{A}{4\pi} N_i(z,p),
\end{eqnarray}
where $E_{\rm kin}/A$ is the energy per nucleon with $E_{\rm kin}=E-m_i c^2$ and mass number $A$.

\subsection{Results}

\begin{figure}
\includegraphics[width=\linewidth]{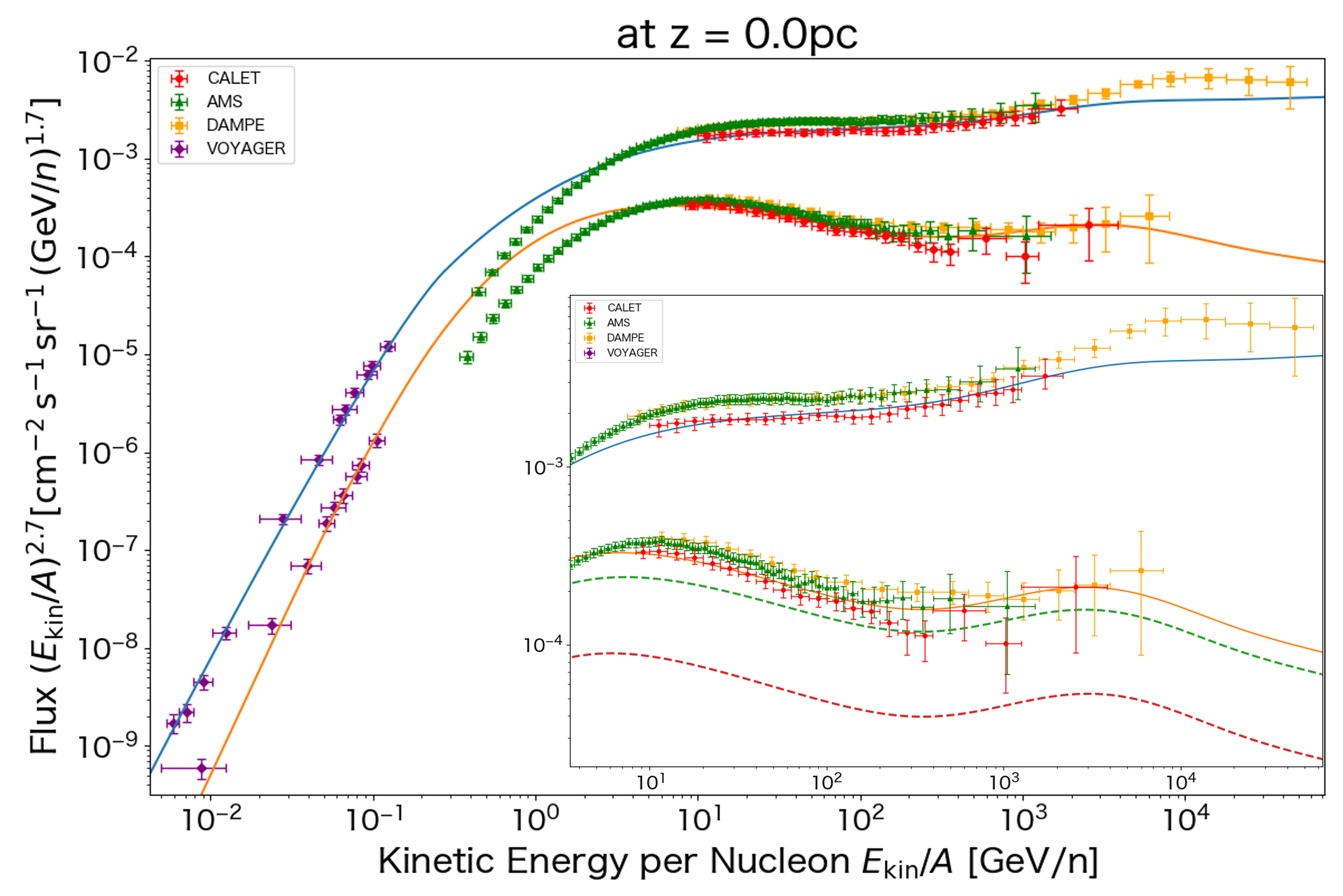}
\caption{CR C and B spectra at $z=0$ and $t=1$ Gyr (solid lines). The data points are from CALET \citep[red,][]{2020PhRvL.125y1102A,2022PhRvL.129y1103A}, AMS-02 \citep[green,][]{2017PhRvL.119y1101A,2018PhRvL.120b1101A}, DAMPE \citep[orange,][]{DAMPE,PhysRevLett.134.191001}, and Voyager \citep[purple,][]{2016ApJ...831...18C}. The inset shows a zoom-up view around the spectral bump. The green and red dashed curves are contributions of $^{11}$B and $^{10}$B, respectively.}
\label{fig:CB}
\end{figure}

As shown in Figure \ref{fig:CB}, our parameter set agrees well with the observed data of the primary and secondary CRs, C and B, respectively. To model low-energy local CRs outside the heliosphere, we have used the data obtained with Voyager \citep{2016ApJ...831...18C}, while the low-energy data of AMS-02 \citep{2017PhRvL.119y1101A,2018PhRvL.120b1101A} are affected by the solar wind.
While several observed datasets are plotted in the figures, we have chosen the parameters that highlight the CALET data points.
As shown in the inset in the figure, the spectral hardenings are reproduced.

\begin{figure*}
\centering
\includegraphics[width=0.8\linewidth]{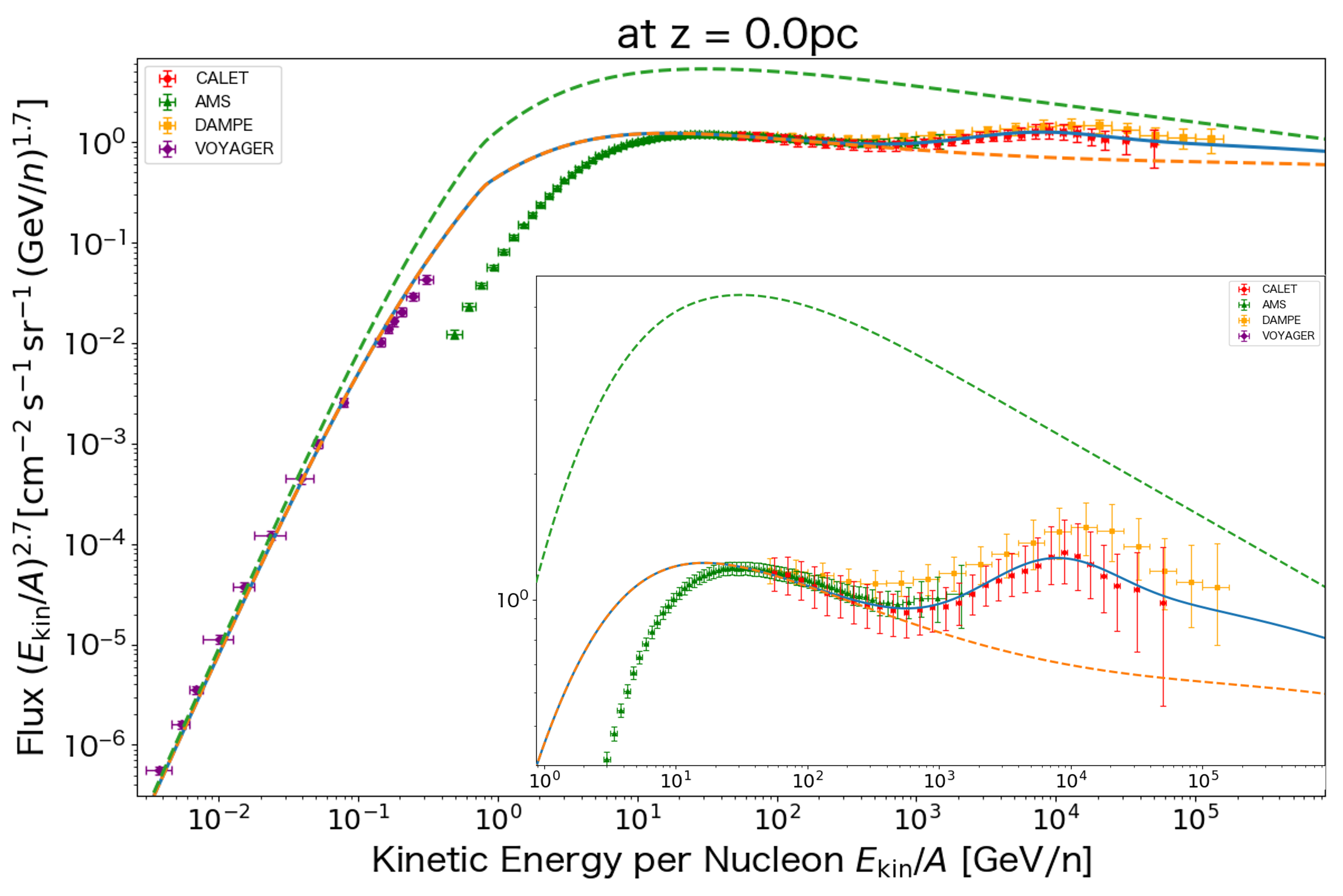}
\caption{CR proton spectrum at $z=0$ and $t=1$ Gyr (blue solid line). The data points are from CALET \citep[red,][]{adriani22}, AMS-02 \citep[green,][]{aguilar15}, DAMPE \citep[orange,][]{DAMPE}, and Voyager \citep[purple,][]{2016ApJ...831...18C}. The inset shows a zoom-up view around the spectral bump. The green dashed curve is the model without the wind, and the orange dashed curve is the model with a constant wind velocity (see Figure \ref{fig:profile}).}
\label{fig:proton}
\end{figure*}

Our main results are shown in Figure \ref{fig:proton}. The sharp hardening followed by softening in the proton spectrum measured with CALET is well reproduced by our model. We also plot the case in which the wind velocity is kept constant beyond the maximum point of equation (\ref{eq:wind}) (see the dashed line in Figure \ref{fig:profile}).
In this case, a slight hardening is observed, but it is not significant enough to support the observed hardening. Therefore, the deceleration of the wind is necessary to reproduce the spectral bump.

\begin{figure}
\includegraphics[width=\linewidth]{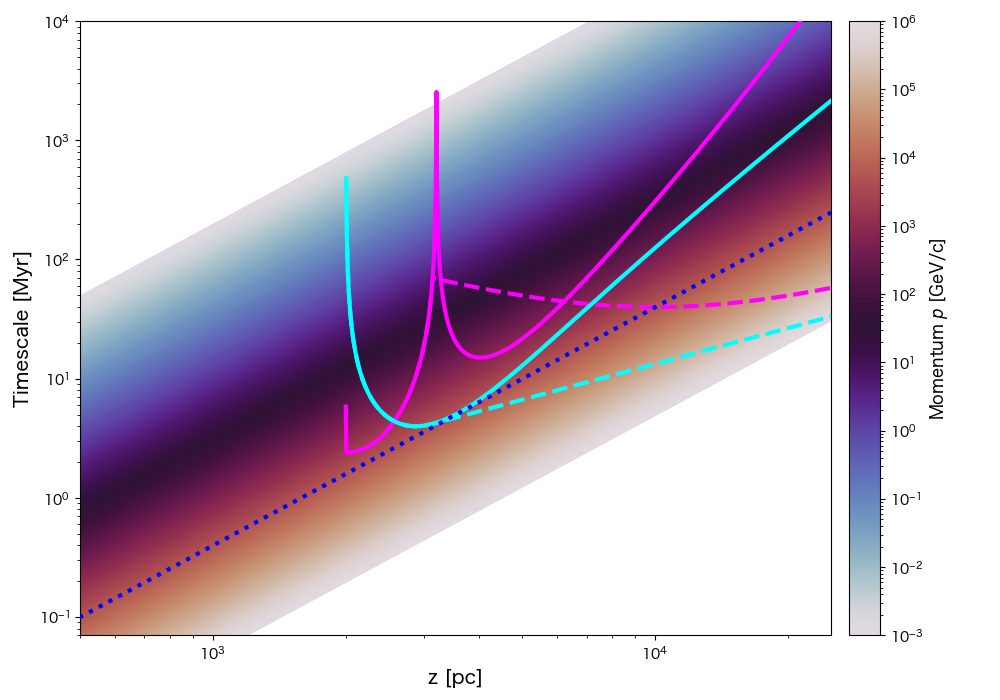}
\caption{Diffusion (color scale) and advection (cyan lines) timescales of protons as a function of the altitude $z$. The blue dotted line is the highlighted diffusion timescale for $p=5\mbox{TeV}/c$. The magenta curves indicate the timescale of adiabatic energy change resulting from the acceleration/deceleration of the wind. The dashed curves correspond to the model with a constant wind velocity (see Figure \ref{fig:profile}).}
\label{fig:time}
\end{figure}

In Figure \ref{fig:time}, we plot the diffusion and wind advection timescales, which are defined as
\begin{eqnarray}
t_{\rm dif}(z,p)=\frac{z^2}{2D(p)},\quad t_{\rm adv}(z)=\frac{z}{V(z)},
\end{eqnarray}
respectively. The latter is independent of momentum $p$.
As shown in the Figure, low-energy CRs diffusively propagate to the wind foot at $z=z_{\rm w}$, beyond which the advection timescale is shorter than the diffusion timescale. For low-energy particles, the wind altitude $z_{\rm w}$ effectively provides the disk size. On the other hand, as shown in Figure \ref{fig:time}, the diffusion timescale is always shorter than the wind advection timescale for protons of $p>5\mbox{TeV}/c$. This threshold momentum roughly corresponds to the peak of the bump in the proton spectrum. Above this energy, the effective size of the disk is $z_{\rm tube}$.
We also plot the timescale of adiabatic energy change
\begin{eqnarray}
t_{\rm adi}=\frac{3}{|\nabla \cdot V(z)|},
\end{eqnarray}
in Figure \ref{fig:time}.
The energy loss in the accelerating wind works for sub-TeV CRs.
However, their long diffusion timescale may prevent them from returning to the disk center before being advected by the wind.
The energy loss due to the wind may be a minor effect for the resultant spectrum at $z=0$.

\begin{figure}
\includegraphics[width=\linewidth]{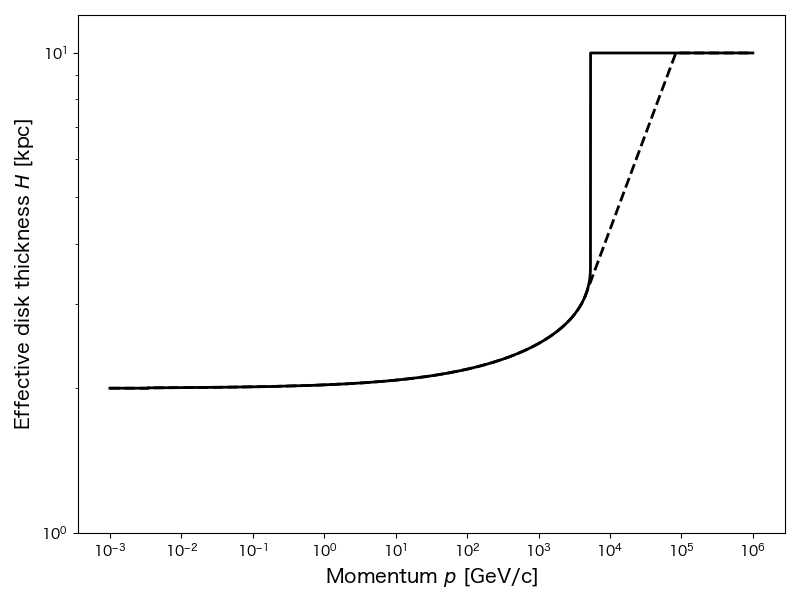}
\caption{Effective disk thickness $H$ as a function of proton momentum. The solid line is our standard case, while the dashed line is the model with a constant wind velocity (see Figure \ref{fig:profile}).}
\label{fig:scale}
\end{figure}

Figure \ref{fig:scale} shows the effective disk thickness $H=\min(H_*,z_{\rm tube})$, where $H_*$ satisfies
\begin{eqnarray}
H_*=\frac{2D(p)}{V(H_*)}.
\end{eqnarray}
At $z=H_*$, the advection and diffusion timescales are the same.
As expected, the thickness is almost $z_{\rm w}=2$ kpc for low-energy CRs, and drastically jumps to $z_{\rm tube}=10$ kpc at $p=5\mbox{TeV}/c$. In the simple leaky box model, in which the volume is proportional to $H$, the CR residence time is $\sim H^2/D$. Then, fixing the diffusion coefficient $D$, the CR density is proportional to $H$.
The low-energy flux ratio of the windless (green dashed) and wind (blue solid) models in Figure \ref{fig:proton} is consistent with the ratio of the effective thickness $z_{\rm tube}/z_{\rm w}=5$. Above $5~\mbox{TeV}/c$ in our fiducial model, the effective thickness increases to 10 kpc, so that the flux level also increases. This is the mechanism of the spectral bump formation in the wind model.
Even above $5~\mbox{TeV}/c$, we cannot perfectly neglect the advection effect. The flux is still lower than that in the windless model.  

\begin{figure}
\includegraphics[width=\linewidth]{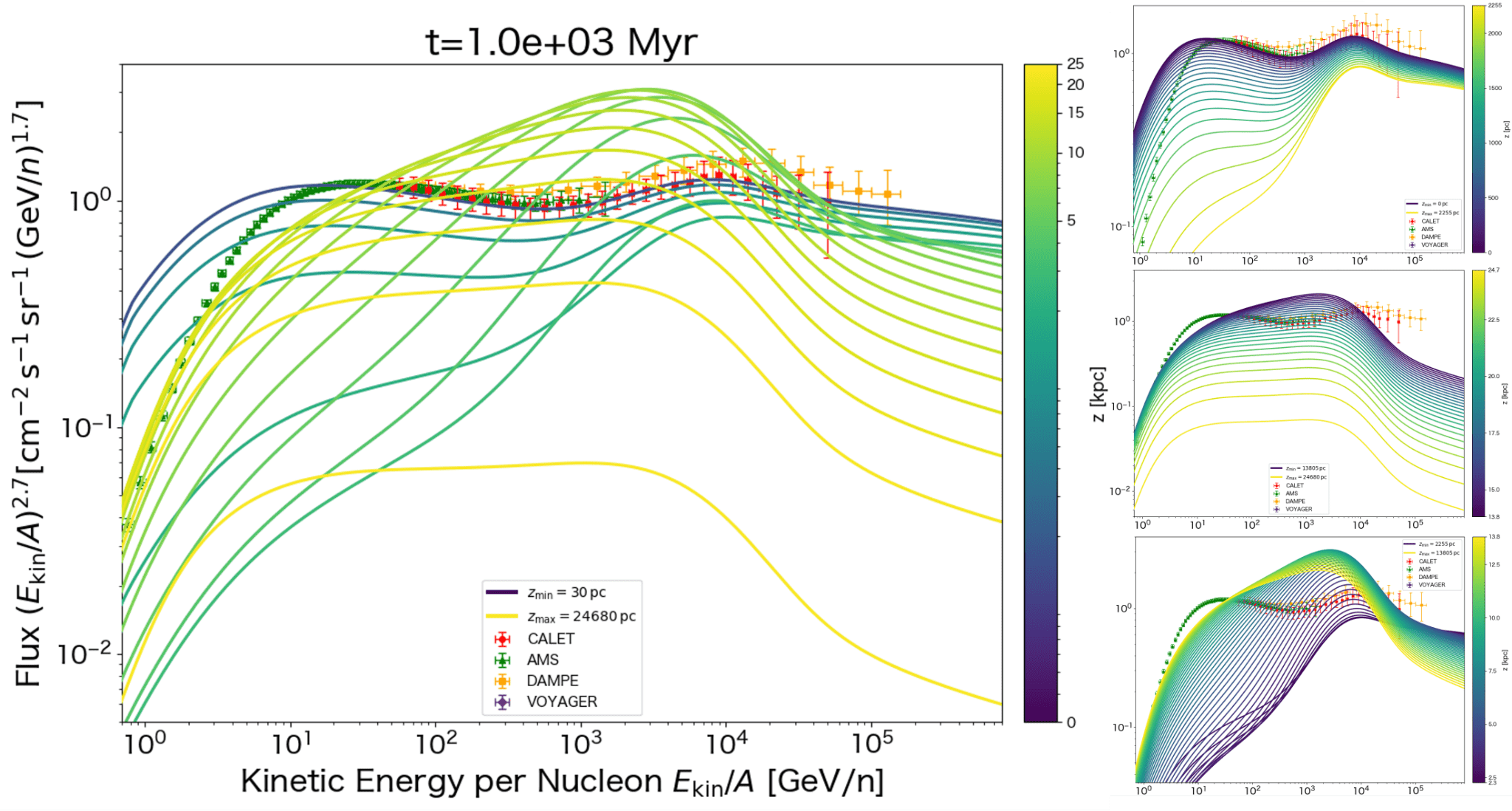}
\caption{Proton spectra at different altitudes from $z=0$ to 25 kpc (left). The right three panels are the same figure divided into three parts based on the altitude ranges, $z=0$ to 2.3 kpc, 2.3 to 13.8 kpc, and 13.8 to 25 kpc, respectively.}
\label{fig:p_z}
\end{figure}

The dashed lines in Figure \ref{fig:time} show the case for the constant wind velocity. In this case, even above $5~\mbox{TeV}/c$, there exists $H_*$, above which the advection is efficient. As indicated in Figure \ref{fig:scale}, the effective box size gradually increases with the CR energy in this case. Therefore, this model cannot produce a sharp hardening in the CR spectra.

\begin{figure}
\includegraphics[width=\linewidth]{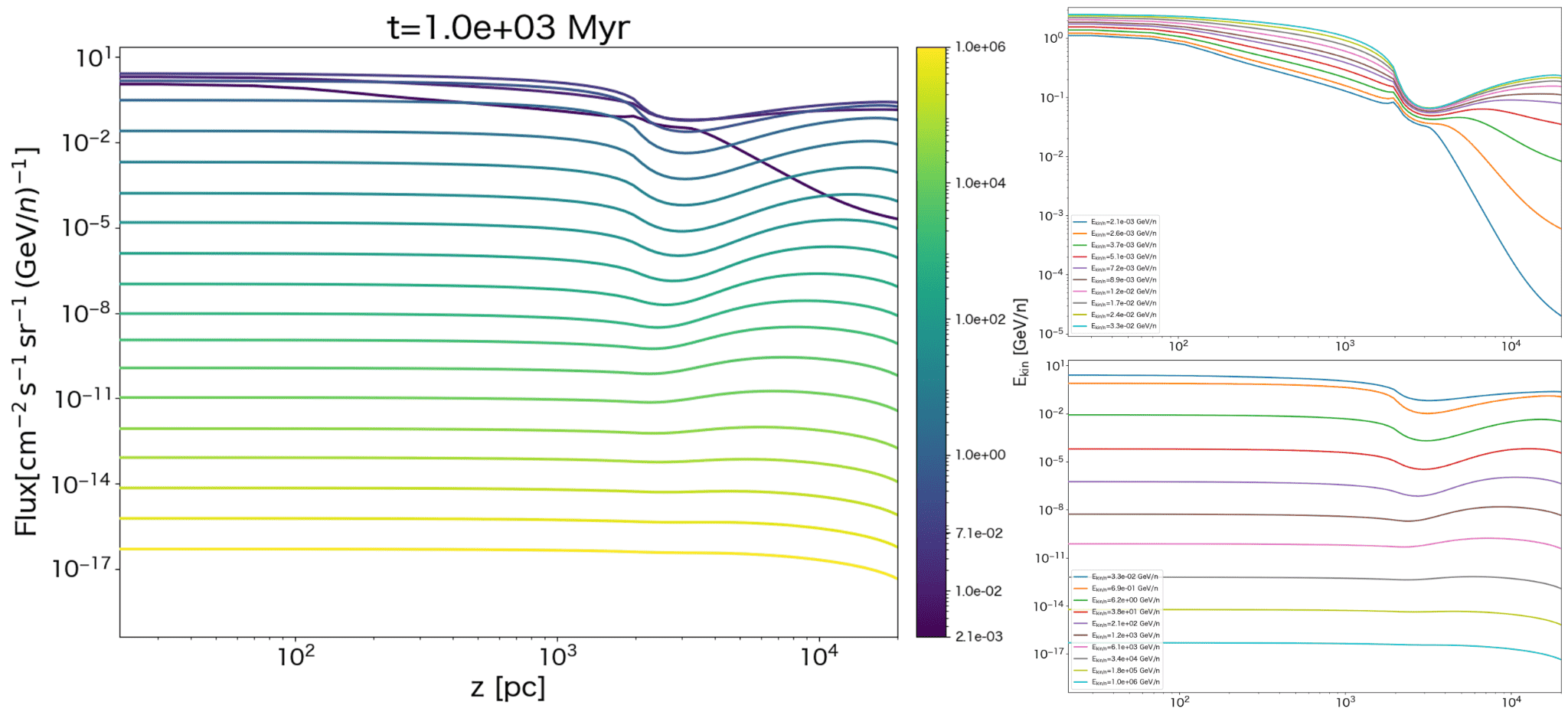}
\caption{CR distributions for different energies from $E_{\rm kin}=2.1 \times 10^{-3}$ GeV to PeV (left). The right two panels are the same figure divided into two parts below and above $E_{\rm kin}=3.3 \times 10^{-2}$ GeV.}
\label{fig:z-p}
\end{figure}

The altitude dependence of the proton spectrum shown in Figure \ref{fig:p_z} is complicated.
For $z<z_{\rm w}$, the flux decreases to achieve the required energy transfer outward via diffusion. In the low-energy region, the energy loss due to collisions further enhances the flux decrease. Above $z_{\rm w}$, the flux below $\sim 5$ TeV turns into an increase. The expansion of the tube cross section leads to a flux decrease again for $z> z_{\rm tube}$.
The spectrum becomes hardest around $z\sim 3$-$5$ kpc, where the spectral index is about $-2.0$.

The CR spatial distributions shown in Figure \ref{fig:z-p} can help interpret the weird flux increase in the wind.
The wind advection effect is prominent for low-energy CRs. 
The wind acceleration and deceleration are clearly reflected in the decrease and increase of the CR density, respectively.
The density increase makes the outward energy transfer via diffusion negative.
In such regions, the energy transfer stagnates.
This is the reason for the CR density peak higher than that at $z=0$.
We can observe that the altitude of the outside peak of the CR flux shifts lower as the CR energy increases.
Figure \ref{fig:time} shows that there is a second $H_*$ for low-energy CRs.
The stagnation point may roughly correspond to this second $H_*$, where the wind advection becomes inefficient again.
This interpretation agrees with the peak shift of the CR density.

\subsection{Discussion}

The wind profile we assumed has reproduced the bump features of the CR spectra.
In our model, the wind starts from a height of $\sim 2$ kpc, accelerates to $\sim 700\mbox{km}~\mbox{s}^{-1}$ with a distance of $d \sim 2$ kpc, and decelerates outside.
The height $z_{\rm w}$ is consistent with the thick disk structure and the scale of the  galactic fountain
\citep[e.g.,][]{shapiro76,lockman84,girichidis18,nakashima18}.
We regard the region below $z_{\rm w}$ as the convection zone, where convection and turbulence flows exist, but the long-term average velocity field may be absent.

The hard CR spectrum around $z>3$ kpc is preferable for the observed hard spectrum in the Fermi bubbles \citep{ackermann14}. CRs escaped from the disk collide with protons in the galactic fountain flow above the Galactic center and produce pions. \citet{shimoda_asano24} show that the $\pi^0$-decay gamma-ray intensity in the central fountain flow is comparable to the observed intensity of the Fermi bubbles without the additional CR injection. While other types of models, including leptonic models, usually require energy injection from nontrivial sources, the Fermi bubble is a by-product of
usual star formation activity in this scenario. In addition, because the hot gas driven by CRs is predicted, the nontrivial bubble morphology, Fermi bubbles inside the eROSITA bubbles, is naturally explained.
In their model, however, reflecting the CR spectrum from the Galactic disk, the resultant gamma-ray spectrum is softer than the observed spectrum.
In our CR advection model, the hard spectral index of $-2$ at the latitude corresponding to the Fermi bubble agrees with the observed spectrum.

Note that the acceleration mechanism of the wind has not been quantitatively established yet. As the radiative cooling in the wind is significant, the gas heating effect by CRs should be included to successfully launch the wind \citep{shimoda22a}.
The required maximum velocity, $\sim 700~\mbox{km}~\mbox{s}^{-1}$, faster than the rotation velocity of the Galactic disk, is not theoretically guaranteed.
Optimistically speaking, outflowing signatures with such a high velocity have been detected in other galaxies \citep[e.g.,][]{2009ApJ...692..187W,2012ApJ...760..127M,2025ApJ...994..102C}.
As will be discussed in the next section, the wind mass-loss rate from the Galactic disk is a few times $M_\odot~ \mbox{yr}^{-1}$. Energetically, the maximum velocity is too fast for a mass-conserving steady outflow with this mass outflow rate.

One point to keep in mind is the fact that the gas in a halo, including the galactic wind, is multiphase \citep[e.g.,][]{2006ApJ...637..333H,2012ARA&A..50..491P}. There are warm dense components and rare hot components.
A fraction of gases in the wind may drop out of the flow and stagnate at a lower altitude as a warm dense component.
The CR advection is regulated by the volume-averaged velocity rather than the mass-weighted velocity,
so that the effective velocity field corresponds to the velocity of the rare hot component, and does not express a mass-conserving steady outflow.
The wind profile proposed in this paper should be regarded as a quasi-steady profile averaged over a significantly long time.
The detection of a very hot gas phase in the circumgalactic medium with a temperature of $\sim 10^7$K \citep{2019ApJ...882L..23D}, which corresponds to $\sim 700~\mbox{km}~\mbox{s}^{-1}$,
may suggest such a high-velocity rare hot component.
A consistent picture of the mixture of multiphase inflows and outflows in the halo \citep{girichidis18,girichidis22,2022MNRAS.517..597C,2025ApJ...987..204S,armillotta25}, including the effects of radiative cooling and CRs with a significant spatial resolution,  remains a challenge that needs to be solved in the future.

The deceleration of the wind is also nontrivial. However, in the Galactic halo, there are stored gases and cold accretion flows. The wind may be gradually mixed with those gases. The wind deceleration profile may be a result of the time-averaged mixing process.

Another possibility is that the required wind profile is not universal in our Galaxy.
The root of the wind ($V=0$) may be nonsteady, hot gases with $10^7$K are intermittently injected, and rarefaction waves propagate.
The required velocity field may be local and temporary with a timescale of $\sim$ Gyr. In this case, the average wind velocity for the whole Galaxy can be as low as $200~\mbox{km}~\mbox{s}^{-1}$.

In this paper, for simplicity, we assume that CRs are strongly coupled with the background gas, so that CRs propagate via only diffusion and advection. However, when the waves that scatter CRs are regulated by CRs themselves, CRs can stream along the magnetic field.
In \citet{2023MNRAS.521.3023T}, where the CR streaming effect is considered in the galactic wind, the effective CR velocity profile is surprisingly similar to ours, with a maximum velocity of $\sim 700~\mbox{km}~\mbox{s}^{-1}$, which is regulated by the Alfv\'en velocity and faster than the gas outflowing velocity of $\sim 200~\mbox{km}~\mbox{s}^{-1}$. If the streaming effect is as significant as their simulation, our velocity profile can be regarded as the streaming velocity rather than the wind velocity.


While we have assumed a simple homogeneous diffusion coefficient, a faster diffusion in the halo than in the disk may be possible.
In this case, the effective escape time from the disk may be maintained even with a lower wind velocity.
So the maximum velocity in this paper is probably not absolute.
However, parameter search involving an increased number of parameters (spatial dependence of $D$) is beyond the scope of this paper.


\section{Gas and Metal Budgets in the Galactic Disk}
\label{sec:bud}

The star formation rate and the metallicity in our Galactic disk have been almost unchanged for the past a few Gyr \citep[e.g.,][]{xiang22}.
Our Galactic disk is in the quasi-equilibrium state.
Especially, the metallicity in the disk should be in an equilibrium state or slowly increasing.
If the present metallicity is decreasing, it should be temporarily enhanced by a recent starburst, which has not been observationally confirmed.
In this context, the galactic wind is an essential factor for the regulation of the star formation rate and gas circulation \citep[e.g.,][]{shimoda22a,shimoda24}.
With our parameters to explain the  CR spectra, we discuss the gas and metal circulation in our galaxy in this section.
The picture in this section is based on \citet{shimoda24}.

As caveats of the following discussion, the metal pollution of the disk by supernova explosion progresses so quickly: the total metal mass ejected by supernovae within $\sim 2$~Gyr is comparable to the total metal mass in the disk gas (see the next subsection). Although the circulation of gas and metals is essential to explain the constant metallicity in the sense of the long term and volume average, we should keep in mind the possibility that our reference of the solar abundance pattern reflects significant short-term and local dispersions.

\subsection{Gas Circulation}
\label{sec:gas-circ}

For simplicity, we divide the region into the disk and the halo.
In the galactic wind model in section \ref{sec:spectra}, we need only the velocity field $V(z)$ to discuss the CR spectra.
So the mass-loss rate via the wind is not determined.
Here, we assume that the gas amount in the disk is in an equilibrium state; the wind mass loss and gas consumption due to star formation balance with the gas supply from the halo.

The CR proton injection rate shown in Table \ref{tab:inj} constrains the star formation rate.
Adopting the effective radius of the Galactic disk as $R_{\rm d}=15$ kpc, the CR proton injection rate in our model is $2.19 \times 10^{48}~\mbox{erg}~\mbox{yr}^{-1}$.
If a supernova releases CRs with 10\% of the explosion energy $10^{51}$ erg and CR protons carry 80\% of the CR energy, the supernova rate is estimated as $2.74\times 10^{-2} \mbox{yr}^{-1}$.
In the standard picture, stars with their initial mass larger than $8 M_\odot$ give rise to a supernova. 
Adopting the initial mass function by \citet{SFR_SNrate}, the star formation rate $\dot{M}_{\rm SF}$ and the supernova rate $\mathfrak{R}_{\rm SN} ~[\mbox{yr}^{-1}]$ are linked as $\dot{M}_{\rm SF}=80 \mathfrak{R}_{\rm SN} M_\odot$. Then, we obtain $\dot{M}_{\rm SF}=2.19 M_\odot \mbox{yr}^{-1}$ from our CR model.

On the other hand, a supernova releases gas into the Galactic disk.
Using the gas amount released from supernovae estimated by \citet{NOMOTO2006}, the initial mass function by \citet{SFR_SNrate} yields the average gas mass (hydrogen and helium) released by a supernova as $10.5 M_\odot$.
Our supernova rate leads to a mass release rate by supernovae as $\dot{M}_{\rm SN}=0.286~ M_\odot \mbox{yr}^{-1}$.

In our gas density model, the hydrogen surface density for $z<2$ kpc is $8.20\times10^6~M_\odot~\mbox{kpc}^{-2}$, which leads to the total gas mass $M_{\rm gas}=5.80\times 10^9~M_\odot$ for our simplified disk with a radius of 15 kpc. If there are no inflows and outflows, the disk gas is consumed in $M_{\rm gas}/(\dot{M}_{\rm SF}-\dot{M}_{\rm SN})\simeq 3$ Gyr.

Cosmological simulations \citep{rodriguez16} suggest that the mass accretion rate from extragalactic space to our Galaxy is $\sim 7 M_\odot \mbox{yr}^{-1}$.
The fresh gas accreted on the galactic halo mixes with the preexisting halo gas. Assuming a quasi-equilibrium state also for the halo gas, the same amount of gas in the halo accretes on the disk. So we adopt the gas accretion rate as $\dot{M}_{\rm AC}=7 M_\odot \mbox{yr}^{-1}$.
Then, the wind mass-loss rate $\dot{M}_{\rm W}$ should satisfy $\dot{M}_{\rm AC}+\dot{M}_{\rm SN}-\dot{M}_{\rm SF}-\dot{M}_{\rm W}=0$.
Finally, we obtain the wind mass-loss rate as $\dot{M}_{\rm W}=5.13 ~M_\odot \mbox{yr}^{-1}$.

The surface density of the wind mass-loss rate is $\dot{M}_{\rm W}/(\pi R_{\rm d}^2)=7.26 \times 10^{-3} M_\odot \mbox{kpc}^{-2} \mbox{yr}^{-1}$.
The wind density along the tube in our model implied from this mass-loss rate is plotted in the upper panel of Figure \ref{fig:profile} as the dashed line for reference.
However, as discussed in the previous section, the velocity field does not necessarily imply a mass-conserving steady outflow ($n_{\rm H} V(z) {\cal A}(z)=$const.).
The effective outflow density may be lower than the plot in Figure \ref{fig:profile}.

Another possible scenario is the slow circulation model. Given $\dot{M}_{\rm SF}-\dot{M}_{\rm SN}=\dot{M}_{\rm AC}-\dot{M}_{\rm W}=1.90~M_\odot \mbox{yr}^{-1}$, we can assume a low wind mass loss such as $\dot{M}_{\rm W}=0.51 ~M_\odot \mbox{yr}^{-1}$, 10\% of the former model, leading to $\dot{M}_{\rm AC}=2.41 ~M_\odot \mbox{yr}^{-1}$. In this model, the halo gas mass continues to increase. Even if all the wind gas gets a high velocity of $V=V_{\rm max}=756 \mbox{km}~\mbox{s}^{-1}$, the implied mass-loss rate ($7.26 \times 10^{-3} M_\odot \mbox{kpc}^{-2} \mbox{yr}^{-1}$)
results in a reasonable value of the surface density of the energy release rate by the wind, $4.13 \times 10^{45} \mbox{erg}~ \mbox{kpc}^{-2} \mbox{yr}^{-1}$, which is comparable to the CR injection rate (see Table \ref{tab:inj}).
However, as will be discussed in the next subsection, the implied low removal rate of metals in the disk causes a problem with the disk metal abundance.

\subsection{Metal Abundance}
\label{sec:metal}

The supernova rate obtained in section \ref{sec:gas-circ} provides the metal supply rate into the Galactic disk gas.
As oxygen nuclei in our galaxy are dominantly produced from supernovae, we discuss $^{16}$O as a representative metal element.
From the supernova model for the metal release of \citet{NOMOTO2006} and the initial mass function by \citet{SFR_SNrate}, the average oxygen mass from a supernova is estimated as $1.23~M_\odot$.
The oxygen injection rate from supernovae is obtained as $\dot{O}_{\rm SN}=1.23~M_\odot \mathfrak{R}_{\rm SN}=3.37\times 10^{-2}M_\odot \mbox{yr}^{-1}$.

Below $z=z_{\rm w}$, the disk gas is assumed to be well mixed by convection and turbulence flows, leading to a homogeneous metallicity.
With the star formation rate obtained in section \ref{sec:gas-circ} and the solar metallicity in Table \ref{table:fi}, the consumption rate of oxygen by star formation is $\dot{O}_{\rm SF}=16 f_{\rm O} \dot{M}_{\rm SF}=2.13 \times 10^{-2}M_\odot \mbox{yr}^{-1}$.
If we consider only the star formation and supernovae, the oxygen gas will double in $16 f_{\rm O} M_{\rm gas}/(\dot{O}_{\rm SN}-\dot{O}_{\rm SF})\simeq 4.5$ Gyr.
Adopting the wind mass-loss rate in section \ref{sec:gas-circ}, the oxygen loss rate by the galactic wind is $\dot{O}_{\rm W}=16 f_{\rm O} \dot{M}_{\rm W}=4.99\times 10^{-2}M_\odot \mbox{yr}^{-1}$.
The quasi-equilibrium oxygen abundance requires
$\dot{O}_{\rm AC}+\dot{O}_{\rm SN}-\dot{O}_{\rm SF}-\dot{O}_{\rm W}=0$,
where $\dot{O}_{\rm AC}$ is the oxygen supply rate by the gas accretion. The above equation requires $\dot{O}_{\rm AC}=3.75\times 10^{-2}M_\odot \mbox{yr}^{-1}$. The mixing of the wind gas and fresh gas accreted from extragalactic space in the halo results in a different metallicity for the gas accreting on the Galactic disk.
Adopting the accretion rate $\dot{M}_{\rm AC}=7 M_\odot \mbox{yr}^{-1}$, the fraction of oxygen in the halo is 0.55 times the solar value. This lower metallicity is consistent with the theoretical model in \citet{shimoda24},
and the observations of HI clouds \citep{ashley24,hayakawa24},
and X-ray emitting gases \citep{miller15} in the halo.

If we consider the slow circulation model, in which $\dot{M}_{\rm W}=0.51 ~M_\odot \mbox{yr}^{-1}$ and $\dot{M}_{\rm AC}=2.41 ~M_\odot \mbox{yr}^{-1}$, the efficiency of the oxygen removal by the wind is very low as $\dot{O}_{\rm W}=4.99\times 10^{-3}M_\odot \mbox{yr}^{-1}$.
In this case, the oxygen amount or abundance cannot be in equilibrium, which requires a negative $\dot{O}_{\rm AC}$.
If the oxygen abundance in the halo is 10\% of the solar abundance, $\dot{O}_{\rm AC}=2.34\times 10^{-3}M_\odot \mbox{yr}^{-1}$, which leads to a total oxygen increase rate of $\dot{O}=9.75\times 10^{-3}M_\odot \mbox{yr}^{-1}$. When we solve the time evolution of $f_{\rm O}$ following $\dot{O}$ as a function of $f_{\rm O}$ and rewind time of $\sim 3$ Gyr, the oxygen abundance becomes zero.
Namely, compared to the steady evolution of our Galaxy during $\sim 10$ Gyr, this rate of increase is too high. We need a more efficient oxygen removal: e.g., a wind mass-loss rate of
$\dot{M}_{\rm W}=1.63 ~M_\odot \mbox{yr}^{-1}$ ($\dot{M}_{\rm AC}=3.53 ~M_\odot \mbox{yr}^{-1}$) to achieve an equilibrium state with an abundance of $0.1 f_{\rm O}$ in the halo.

\subsection{{\rm Be} Abundance}
\label{sec:be}

Here, we discuss elements synthesized via CR spallation. While the representative element is boron, 
$^{11}$B can be produced from supernovae by the $\nu$-process \citep{1990ApJ...356..272W,2005PhLB..606..258H,2012A&A...542A..67P,2018ApJ...865..143S} too. So we consider beryllium, whose production from supernovae is negligible.

There are two processes to produce $^9$Be in the gas of the Galactic disk.
The first one is the spallation between CR protons and C/N/O in the gas. Adopting the CR proton density $N_{\rm p}(z,p)$, the Be production rate is
\begin{eqnarray}
\dot{n}_{\rm Be,1}(z)=
\sum_{i}
\int dp \sigma_{i{\rm Be}}(p)\,
v(p)\,
N_{\rm p}(z,p) f_i n_\mathrm{H}(z),
\end{eqnarray}
where $i=$C, N, and O.
The second one is the spallation between CR C/N/O and protons in the gas. Low-energy secondary Be CRs lose their energy via collision with the background gas before escaping from the disk.
Using the CR Be injection rate $Q_{\rm Be}(z,p)$ in equation (\ref{eq:sec}), the Be production rate via this process is calculated as
\begin{eqnarray}
\dot{n}_{\rm Be,2}(z)=
\int dp Q_{\rm Be}(z,p) \exp
\left( -\frac{\tau_{\rm Be,c}(z,p)}{\tau_{\rm dif}(p)} \right),
\end{eqnarray}
where the cooling time of Be CRs $\tau_{\rm Be,c}(z,p)$
is defined with $\dot{p}_{\rm Be}$ in equation (\ref{eq:pdot}) as
\begin{eqnarray}
\tau_{\rm Be,c}(z,p)=
\int_p^0 dp \frac{1}{\dot{p}_{\rm Be}(p,z)},
\end{eqnarray}
and the diffusion time $\tau_{\rm dif}(p)$ is defined with the effective disk thickness $H_*$ as
\begin{eqnarray}
\tau_{\rm dif}(p)=
\frac{H_*^2}{2D(p)}.
\end{eqnarray}
Integrating the values between $-2$ and 2 kpc, we obtain the surface production rate $\dot{\Sigma}_{\rm Be,1}=1.92 \times 10^{43}~\mbox{kpc}^{-2} \mbox{yr}^{-1}$ and $\dot{\Sigma}_{\rm Be,2}=1.12 \times 10^{43}~\mbox{kpc}^{-2} \mbox{yr}^{-1}$, respectively.
The total Be production rate in the disk of $R_{\rm d}=15$ kpc is
$\dot{Be}_{\rm CR}=1.62\times 10^{-10}M_\odot \mbox{yr}^{-1}$.

Similarly to the oxygen case, the consumption rate by star formation is calculated as $\dot{Be}_{\rm SF}=4.65\times 10^{-10}M_\odot \mbox{yr}^{-1}$.
First, we consider the quasi-equilibrium model of $\dot{M}_{\rm W}=5.13 ~M_\odot \mbox{yr}^{-1}$. The Be loss rate by the wind is $\dot{Be}_{\rm W}=1.09\times 10^{-9}M_\odot \mbox{yr}^{-1}$.
If the Be abundance in the halo is 0.55 times the solar value, which is the required abundance for the oxygen equilibrium, the Be accretion rate from the halo is $\dot{Be}_{\rm AC}=8.19\times 10^{-10}M_\odot \mbox{yr}^{-1}$ for $\dot{M}_{\rm AC}=7 M_\odot \mbox{yr}^{-1}$.
The total Be increase rate becomes negative as $\dot{Be}=\dot{Be}_{\rm CR}+\dot{Be}_{\rm AC}-\dot{Be}_{\rm SF}-\dot{Be}_{\rm W}=-5.7 \times 10^{-10}M_\odot \mbox{yr}^{-1}$.
This decrease of Be in the present Galactic disk is an irrational result. To achieve an equilibrium for the Be abundance, the Be abundance in the halo must be 0.94 times the solar value.
Though this value is lower than the disk abundance, the Be/O ratio in the halo is higher than the solar value.
As the seeds of Be nuclei are O or C, the Be nuclei in the halo were produced in the past with a higher efficiency than that in the present Galactic disk.
This scenario may imply that Be nuclei in the halo are fossil nuclei, indicating a higher CR density in the past.

Finally, let us consider the slow circulation model ($\dot{M}_{\rm W}=1.63 ~M_\odot \mbox{yr}^{-1}$, $\dot{M}_{\rm AC}=3.53 ~M_\odot \mbox{yr}^{-1}$, and 10\% of the solar abundance in the halo). This also results in a negative value of $\dot{Be}$ as $-5.7 \times 10^{-10}M_\odot \mbox{yr}^{-1}$.
To make $\dot{Be}$ positive, we need more than 86\% of the solar abundance in the halo, even in this model.

The Be problem is due to the high removal rate by star formation.
The caveats are uncertainties in the Be abundance and the estimate of the star formation rate. The abundance may be locally or temporarily enhanced around the Sun. If the average $f_{\rm Be}$ or the supernova rate (equivalently, the star formation rate) is lower than the present local value, the contradiction in the Be production can be relaxed,
although the supernova rate we used is rather lower than that in other models \citep[see e.g.,][]{2019PhRvD..99j3023E}.
The uncertainty in the local gas density can lead to an underestimation of the Be production rate.
As the observation constrains the Be CR flux, the Be injection from C/N/O CRs ($\dot{\Sigma}_{\rm Be,2}$) is not so altered by the increase of the local gas density.
If we increase the gas density by a factor of four, the higher $\dot{\Sigma}_{\rm Be,1}$ makes the Be injection rate equal to the consumption rate by the star formation.

\section{Summary}
\label{sec:summary}

We have demonstrated that the advection effect by the galactic wind can reproduce the bump structures at ${\cal R}\sim$ TV in the CR spectra without introducing a break in the power-law dependence of the diffusion coefficient.
The required velocity field of the wind drastically changes the effective box size at ${\cal R}\sim$ TV, at which the advection and diffusion timescales are comparable.
This implies a rather high wind velocity of $700~\mbox{km}~ \mbox{s}^{-1}$.
Although the required high velocity should be examined observationally and theoretically in the future, we suppose that only a small fraction of the wind gas is responsible for this high-velocity field.

Our model predicts a hard CR spectrum below $\sim$ TeV with an index of $\sim 2$ at an altitude of $\sim 3$-5 kpc.
With this hard CR spectrum, the gamma-ray spectrum of the Fermi bubbles can be explained by $\pi^0$-decay emission due to collisions between halo CRs and cold accretion flows \citep{shimoda_asano24}.

Based on the obtained CR fluxes, we have discussed the matter circulation in our Galaxy with the wind. As discussed in \citet{shimoda24}, the metal removal from the disk by the wind is necessary to maintain the metal abundance steadily during a few Gyr.
However, even with the model of the wind and accretion consistent with the oxygen abundance, the production rate of beryllium, which originates from CR spallation, is so low for maintaining the Be abundance. The consumption rate by star formation is higher than the production rate, though we should note the uncertainties in the parameter estimates.
To compensate Be nuclei by the gas accreted from the halo, the ratio Be/O in the halo should be larger than that in the disk gas.
If this is the case, this is a smoking gun of the efficient synthesis of Be nuclei in the past starburst era, when the CR density was higher than the present value.

\begin{acknowledgments}

We gratefully acknowledge the support from the
CALET Collaboration team.
The authors also thank R. Sawada for the useful suggestions.
We also thank the anonymous referee for the helpful comments.
This work is supported by the joint research program of the Institute for Cosmic Ray Research (ICRR), the University of Tokyo, and KAKENHI grant Nos. 23H04899, 24H00025, and 25K07352 (K.A.), 24K00677, and 25H00394 (J.S.).

\end{acknowledgments}

\appendix
\section{CR Spectra for other Nuclides} \label{sec:app1}

\begin{figure}
\includegraphics[width=\linewidth]{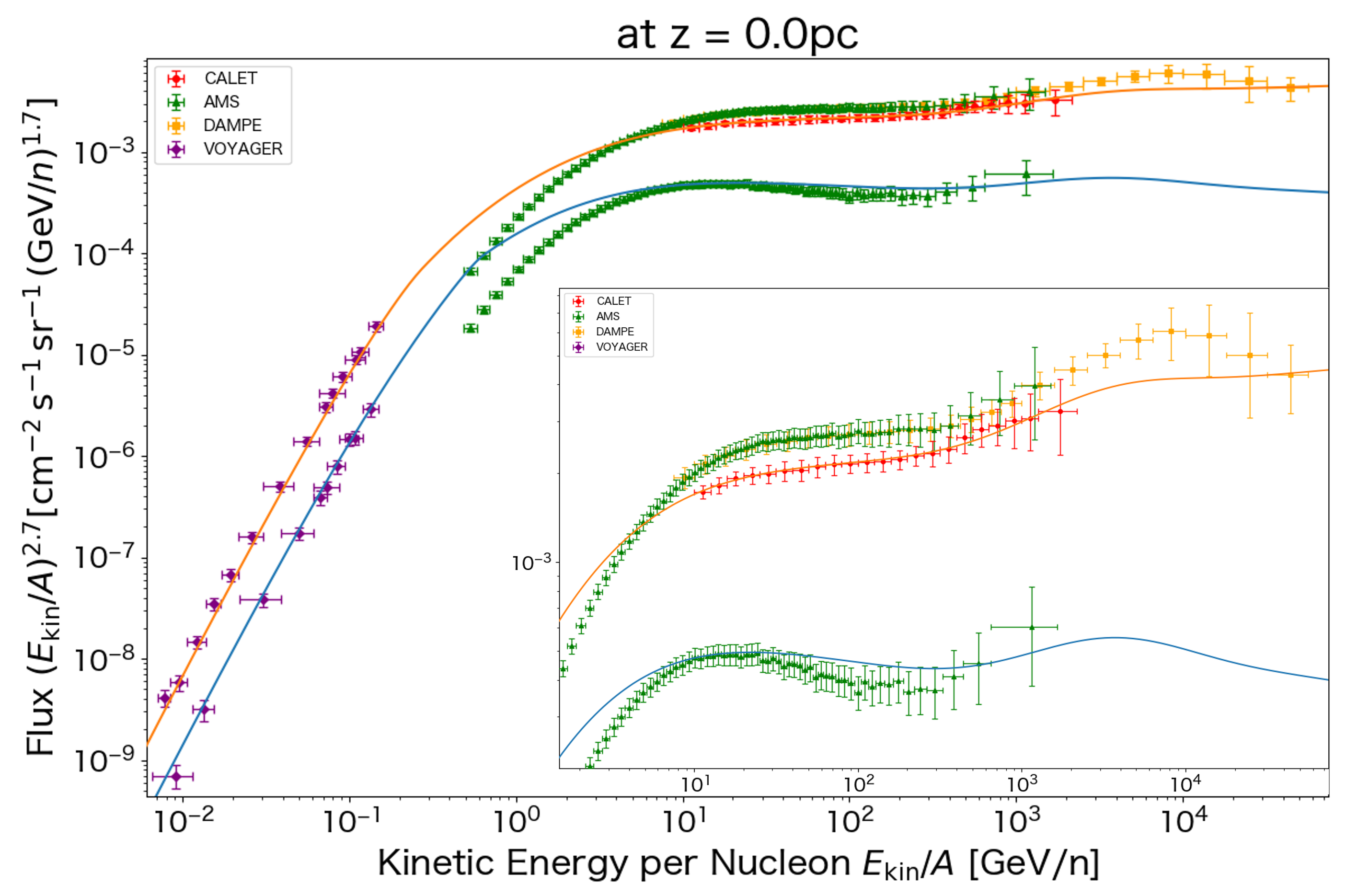}
\caption{CR O and N spectra at $z=0$ and $t=1$ Gyr (solid lines). The data points are from CALET \citep[red,][]{2020PhRvL.125y1102A}, AMS-02 \citep[green,][]{2017PhRvL.119y1101A,PhysRevLett.121.051103}, DAMPE, and Voyager \citep[purple,][]{2016ApJ...831...18C}. The inset shows a zoom-up view around the spectral bump.}
\label{fig:ON}
\end{figure}

\begin{figure}
\includegraphics[width=\linewidth]{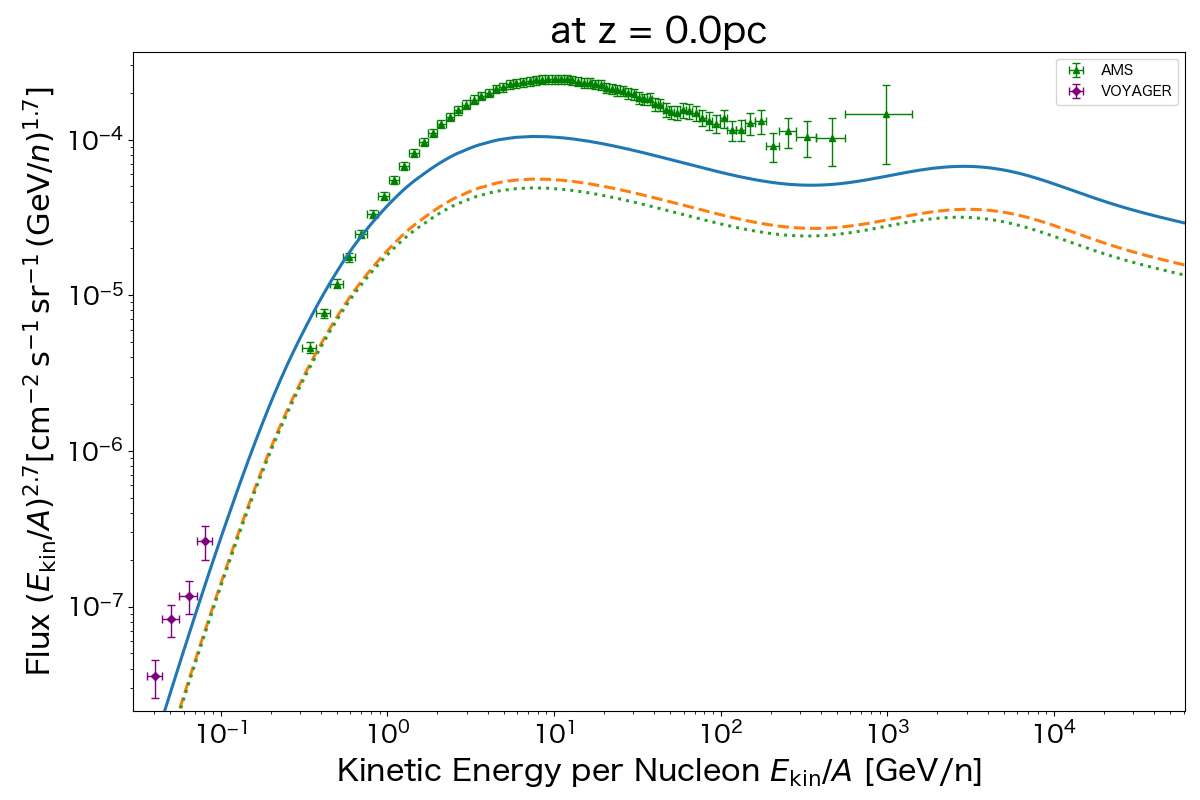}
\caption{Total CR Li spectrum at $z=0$ and $t=1$ Gyr (blue solid line). The data points are from AMS-02 \citep[green,][]{2018PhRvL.120b1101A}, and Voyager \citep[purple,][]{2016ApJ...831...18C}. The orange dashed and green dotted lines are for $^6$Li and $^7$Li, respectively.}
\label{fig:Li}
\end{figure}

\begin{figure}
\includegraphics[width=\linewidth]{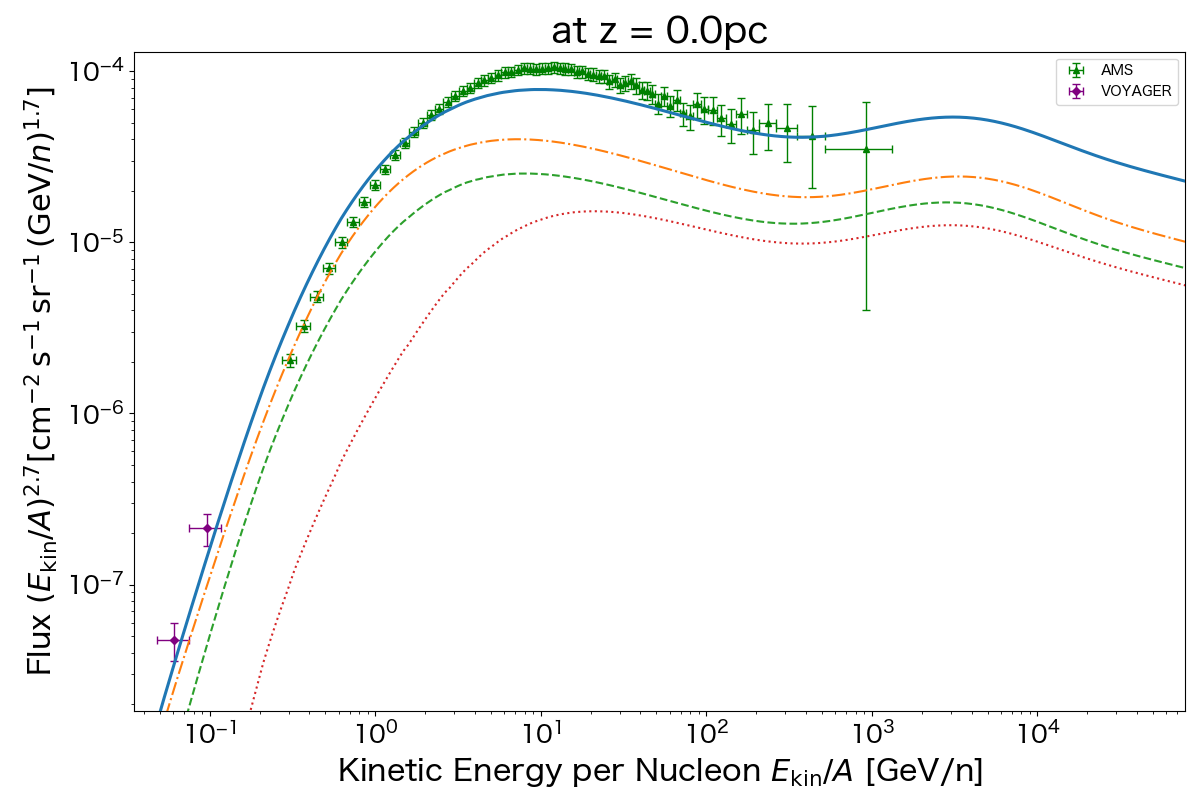}
\caption{Total CR Be spectrum at $z=0$ and $t=1$ Gyr (blue solid line). The data points are from AMS-02 \citep[green,][]{2018PhRvL.120b1101A}, and Voyager \citep[purple,][]{2016ApJ...831...18C}. The orange dashed-dotted, green dotted, and magenta dotted lines are for $^7$Be, $^9$Be, and $^{10}$Be, respectively.}
\label{fig:Be}
\end{figure}

We show spectra for other kinds of CR nuclei. Figure \ref{fig:ON} shows primary CRs of O and N.
Figures \ref{fig:Li} and \ref{fig:Be} show secondary CRs of Li and Be, respectively. As we have chosen the values of the injection parameters to reproduce the B spectrum intensively, given the cross sections for Li and Be, the model spectra for Li and Be deviate from the data.
This mismatch may be due to the uncertainty in the cross sections.
However, the deviation in the Be spectrum is not so large that the discussion concerning the Be production in the main text does not need to be changed.
The ratio $^{10}$Be/$^9$Be at 1 GeV/n in our model is 0.14, which is close to the value measured with AMS-02 \citep{Dimiccoli:2023/s}.

\bibliography{reference_paper}{}
\bibliographystyle{aasjournalv7}

\end{document}